\documentclass[fleqn,usenatbib]{mnras}

\usepackage{newtxtext,newtxmath}

\usepackage[T1]{fontenc}

\DeclareRobustCommand{\VAN}[3]{#2}
\let\VANthebibliography\thebibliography
\def\thebibliography{\DeclareRobustCommand{\VAN}[3]{##3}\VANthebibliography}

\usepackage{graphicx}	
\usepackage{amsmath}	
\usepackage[switch,mathlines]{lineno}

\newcommand{\AGN}{\mathrm{AGN}}
\newcommand{\galaxy}{\mathrm{galaxy}}
\newcommand{\hMpc}{h^{-1}\mathrm{Mpc}}
\newcommand{\hMsun}{h^{-1}\mathrm{M_\odot}}
\newcommand{\Msun}{\mathrm{M_\odot}}
\newcommand{\halo}{\mathrm{halo}}
\newcommand{\duty}{\mathrm{duty}}
\newcommand{\CCF}{\mathrm{CCF}}
\newcommand{\ACF}{\mathrm{ACF}}

\title[Clustering analysis of JWST AGNs]{The nature of low-luminosity AGNs discovered by JWST based on clustering analysis: progenitors of low-$z$ quasars?}

\author[J. Arita et al.]{
Junya Arita,$^{1}$\thanks{E-mail: jarita@astron.s.u-tokyo.ac.jp}
Nobunari Kashikawa, $^{1,2}$
Masafusa Onoue, $^{3,4,5}$
Takehiro Yoshioka, $^{1}$
Yoshihiro Takeda, $^{1}$
\newauthor
Hiroki Hoshi, $^{1}$
and Shunta Shimizu $^{1}$
\\
$^{1}$Department of Astronomy, School of Science, The University of Tokyo, 7-3-1, Hongo, Bunkyo-ku, Tokyo, 113-0033, Japan\\
$^{2}$Research Center for the Early Universe, The University of Tokyo, 7-3-1 Hongo, Bunkyo-ku, Tokyo, 113-0033, Japan\\
$^{3}$Kavli Institute for the Physics and Mathematics of the Universe (WPI), The University of Tokyo Institutes for
Advanced Study, The University of Tokyo, Kashiwa, \\
Chiba 277-8583, Japan\\
$^{4}$Kavli Institute for Astronomy and Astrophysics, Peking University, Beijing 100871, China\\
$^{5}$Center for Data-Driven Discovery, Kavli IPMU (WPI), UTIAS, The University of Tokyo, Kashiwa, Chiba 277-8583, Japan
}

\date{Accepted XXX. Received YYY; in original form ZZZ}

\pubyear{\the\year{}}

\begin{document}
\label{firstpage}
\pagerange{\pageref{firstpage}--\pageref{lastpage}}
\maketitle

\begin{abstract}
James Webb Space Telescope (JWST) has discovered many faint AGNs at high-$z$ by detecting their broad Balmer lines. 
However, their high number density, lack of X-ray emission, and overly high black hole masses with respect to their host stellar masses suggest that they are a distinct population from general type-1 quasars.
Here, we present clustering analysis of 27 low-luminosity broad-line AGNs found by JWST (JWST AGNs) at $5<z<6$ based on cross-correlation analysis with 679 photometrically-selected galaxies to characterize their host dark matter halo (DMH) masses.
From the angular and projected cross-correlation functions, we find that their typical DMH mass is $\log (M_\halo/\hMsun) = 11.46_{-0.25}^{+0.19},$ and $11.53_{-0.20}^{+0.15}$, respectively.
This result implies that the host DMHs of these AGNs are $\sim1$ dex smaller than those of luminous quasars.
The DMHs of the JWST AGNs at $5<z<6$ are predicted to grow to $10^{12\,\mathchar`-13}\,\hMsun$ at $z\lesssim3$, which is comparable to that of a more luminous quasar at the same epoch.
Applying the empirical stellar-to-halo mass ratio to the measured DMH mass, we evaluate their host stellar mass as $\log(M_*/\Msun)=9.48_{-0.41}^{+0.31},$ and $9.60_{-0.33}^{+0.24}$, which are higher than some of those estimated by the SED fitting.
We also evaluate their duty cycle as $f_\duty=0.37_{-0.15}^{+0.19}$ per cent, corresponding to $\sim4\times10^6$ yr as the lifetime of the JWST AGNs.
While we cannot exclude the possibility that the JWST AGNs are simply low-mass type-1 quasars, these results suggest that the JWST AGNs are a different population from type-1 quasars and the progenitors of quasars at $z\lesssim3$.
\end{abstract}




\section{Introduction}
Active Galactic Nuclei (AGNs) are the extremely bright population in the Universe.
Driven by the central supermassive black holes (SMBHs) \citep{Kormendy1995}, the AGNs outshine their host galaxies by releasing part of the gravitational potential of accreting massive gas and stars onto the accretion disks as radiation energy \citep{Salpeter1964, Lynden-Bell1969}.
Previous measurements suggest a tight relationship between the SMBH mass and the host bulge or stellar mass \citep{Kormendy2013}, implying co-evolution between SMBHs and their host galaxies.
Unveiling the co-evolution mechanism is one of the greatest goals of modern galactic astronomy.

Recently, deep IR observation by the James Webb Space Telescope (JWST) discovered many low-luminosity objects with broad Balmer lines with $\mathrm{FWHM}\gtrsim 1000\,\mathrm{km\,s^{-1}}$ at $4<z<7$ (e.g. \citealp{Kocevski2023, Maiolino2023, Harikane2023, Greene2024, Matthee2024, Kocevski2024, Taylor2024}).
The origin of the broad lines for the objects is under debate, but they are usually believed to originate in the broad line regions in AGNs; therefore we hereafter call these objects JWST AGNs.
The high sensitivity of JWST allows the study of AGNs that are much fainter ($M_{UV}\sim-17$) and have been unexplored by ground-based telescope observations.
In addition, JWST discovers remarkable objects named little red dots (LRDs), one of the populations in the JWST AGNs, which are characterized by blue UV excess and red optical slope, and compact morphology in the rest-frame optical images \citep{Kocevski2023, Kokorev2023, Barro2024, Greene2024, Matthee2024, Kocevski2024, Wang2024, Akins2024}.

The observed number density of the JWST AGNs is $>1\mathchar`-2$ dex higher than the extrapolation to the faint-end of the quasar luminosity functions (LFs) \citep{Maiolino2023, Harikane2023, Kocevski2024}.
If their escape fraction of ionizing photons is as high as that of quasars, the JWST AGNs may play a non-negligible role in the reionization (e.g. \citealp{Giallongo2015, Finkelstein2019, Giallongo2019, Boutsia2021, Grazian2022}).
In that case, a modification would be required to the current prevailing reionization scenario, in which star-forming galaxies are the main contributors \citep{Robertson2015}.
In addition, the JWST AGNs display some unfamiliar features.
\citet{Padmanabhan2023} argued that the X-ray background would be inconsistent with the current observation if the overabundant JWST AGNs emit X-rays as strong as quasars, which implies that the JWST AGNs are a distinct AGN population from normal quasars.
\citet{Yue2024} obtained tentative detections from the stacked X-ray images of 34 spectroscopically confirmed LRDs in both soft (0.5\--2 keV) and hard bands (2\--8 keV) with $2.9\sigma$ and $3.2\sigma$ significance, respectively, although the empirical relation with H$\alpha$ luminosity suggests clear detection of X-ray emission.
\citet{Maiolino2024} also reported that the majority of the 71 JWST AGNs at $2<z<11$ were not detected by the Chandra observations.
\citet{Maiolino2023} found that most of the JWST AGNs have significantly overmassive SMBHs compared with their host stellar mass.
The deviation from the local relation (e.g. \citealp{Reines2015}) is difficult to explain only by the selection effect.
\citet{Perez-Gonzalez2024} suggested from MIRI observations that many of the observed LRDs may be extremely intense and compact starburst galaxies based on the best-fitting spectral energy distribution (SED) models.
\citet{Kokubo2024} reported, based on multi-epoch photometry of five broad H$\alpha$ emitters and LRDs, no time variability of AGNs with $M_\mathrm{BH}\sim10^7\,\Msun$, even though their typical timescales are shorter than the sampling interval.
They also suggest that the origin of the broad lines other than AGNs may be unusually fast outflows or Raman scattering of stellar UV continua.
This situation strongly recommends that we should elucidate the nature of the JWST AGNs.

While multiwavelength observations are effective methods to explore the nature of the JWST AGNs, clustering analysis is also useful in differentiating them from quasars.
Clustering analysis can reveal the typical dark matter halo (DMH) mass of the objects. 
The gravitational potential of DMHs plays an important role in accumulating the gas, which is consumed to form stars; hence more massive DMHs can harbour galaxies with more massive stellar mass \citep{White1978}.
Referring to the DMH mass function for the DMH mass range of quasars derived by clustering analysis deduces the number density of entire SMBHs in the Universe.
Comparing the number density with that of quasars estimated by the LF derives the fraction of active SMBHs, which can be regarded as a duty cycle.
These physical quantities, especially in the early Universe, are key to understanding how the co-evolution is constructed and how SMBHs with $M_\mathrm{BH}\gtrsim10^8\,\mathrm{M_\odot}$ are formed as the high-$z$ quasars at $z\gtrsim7$ (e.g. \citealp{Mortlock2011, Banados2018, Matsuoka2019, Yang2020}) pose a challenge to theoretical models on their rapid growth.

Recently, it has become possible to evaluate the auto-correlation function of quasars even at $z\sim6$ owing to the high sensitivity and the large field of view of Hyper Suprime-Cam mounted on the Subaru Telescope.
\citet{Arita2023} used 107 quasars spectroscopically identified in the Subaru High-$z$ Exploration of Low-Luminosity Quasars (SHELLQs; \citealp{SHELLQs1}) and reported that the typical DMH mass of quasars at $z\sim6$ is $\log (M_\halo/\hMsun)=12.7_{-0.7}^{+0.4}$.
They found that the typical DMH mass of type-1 quasars does not change over cosmic time.
They also deduced the host stellar mass of quasars as $\log (M_*/\mathrm{M_\odot})=10.97_{-0.70}^{+0.39}$ assuming the empirical relation between the DMH mass and the stellar mass \citep{Behroozi2019}.
\citet{Eilers2024} performed cross-correlation analysis with four bright quasars and surrounding [O \textsc{iii}] emitters by JWST NIRCam's slitless spectroscopy (Emission-line galaxies and Intergalactic Gas in the Epoch of Reionization, EIGER; \citealp{Kashino2023}) and estimated the minimum DMH mass to host a quasar as $\log (M_{\halo,\mathrm{min}}/\Msun)=12.30\pm0.14$. 
They also estimated the duty cycle as $f_\duty=0.08_{-0.06}^{+0.17}$ per cent.

In this paper, we perform the clustering analysis of the JWST AGNs to unveil their DMH mass.
Although many AGN candidates have already been reported in the JWST public data (e.g.\citealp{kokorev2024, Kocevski2024, Akins2024}) and the number of spectroscopically confirmed AGNs is increasing (e.g.\citealp{Maiolino2023, Harikane2023, Greene2024, Matthee2024, Kocevski2024, Taylor2024}), the survey area is still small and their surface number density is not yet sufficient to evaluate their auto-correlation signal.
Therefore, in order to obtain more robust signals, we perform cross-correlation analysis with the JWST AGNs and surrounding galaxies.
Based on the comparison of the DMH mass between the JWST AGNs and quasars, we discuss whether the JWST AGNs and quasars are the same populations.
In addition, we also calculate the host stellar mass and duty cycle of the JWST AGNs and track $M_\mathrm{BH}/M_*$, which can provide additional information on the nature of the JWST AGNs.

The structure of this paper is as follows.
We remark on the JWST AGN and galaxy sample and their selection to evaluate the correlation functions in Section \ref{sec:data}.
Section \ref{sec:clustering_analysis} presents the details of the clustering analysis based on the angular correlation function (Section \ref{subsec:angular}) and the projected correlation function (Section \ref{subsec:projected}).
In Section \ref{sec:discussion}, we show our results and compare them with the previous studies executing the clustering analysis with quasars.
We summarise our results and conclude in Section \ref{sec:summary}.
We adopt flat $\Lambda$ cold dark matter cosmology with $h=0.7$, $\Omega_m=0.3$, $\Omega_\lambda=0.7$, and $\sigma_8=0.81$ through this paper.
All magnitudes in this paper are presented in the AB system \citep{Oke1983}.

\section{Data and Sample Selection} \label{sec:data}

\subsection{JWST AGNs}
\label{subsec:jwst_agn} 
\begin{figure}
    \centering
    \includegraphics[width=\columnwidth]{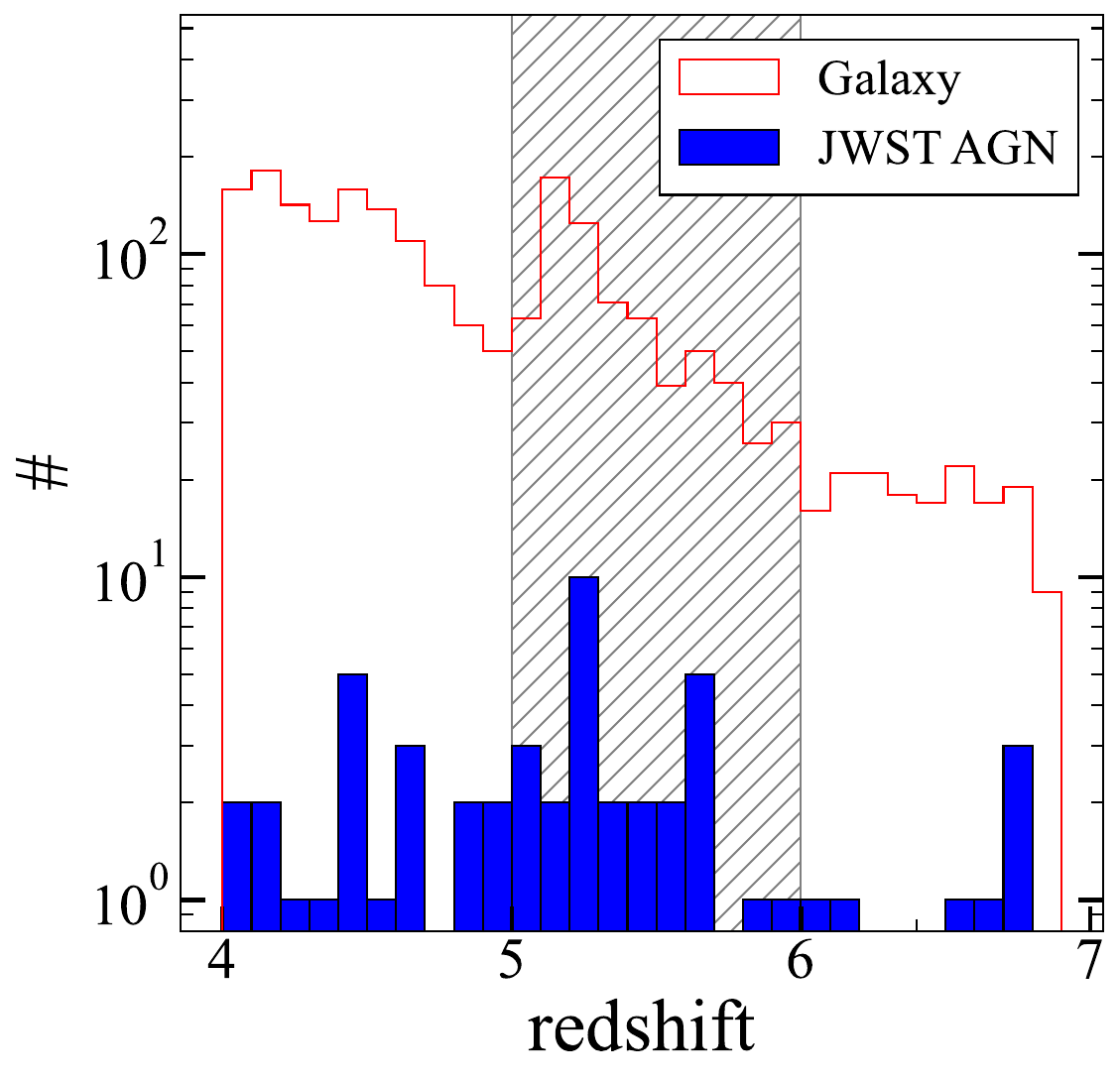}
    \caption{The redshift distribution of the spectroscopic redshifts of the JWST AGNs (blue) and the photometric redshifts of the galaxies (red) samples based on the literature \citep{Maiolino2023, Harikane2023, Kocevski2024, Matthee2024, Taylor2024}.
    The hatched region shows the galaxy and AGN samples used in this analysis.}
    \label{fig:sample_histogram}
\end{figure}

We select JWST AGNs by compiling the literature on spectroscopic observation of AGNs with NIRSpec or NIRCam grism \citep{Maiolino2023, Harikane2023, Kocevski2024, Matthee2024, Taylor2024} identified in the following public fields: the JWST Advanced Deep Extragalactic Survey (JADES; \citealp{Eisenstein2023}); First Reionization Epoch Spectroscopically Complete Observations (FRESCO; \citealp{Oesch2023}); Cosmic Evolution Early Release Science Survey (CEERS; \citealp{Finkelstein2023}); Public Release IMaging for Extragalactic Research (PRIMER) survey \citep{Dunlop2021}\footnote{We only use UDS field in this analysis.}; Red Unknowns: Bright Infrared Extragalactic Survey (RUBIES; \citealp{de_Graaff2024, Wang2024_RUBIES}).
We note that some of the JWST AGNs in the CEERS field are excluded because they are located in the region where only the NIRSpec observation is performed, hence there are no catalogued galaxies with NIRCam photometry around them.
We only use spectroscopically confirmed AGNs with broad Balmer line components with $\mathrm{FWHM}\gtrsim1000\,\mathrm{km\,s^{-1}}$, i.e., the same feature as type-1 AGNs.
AGNs with only narrow Balmer lines that do not meet the above conditions are excluded here because the clustering strength of obscured AGNs may differ from that of type-1 AGNs (e.g. \citealp{Hickox2011}).
No limits are placed on the luminosity of the JWST AGNs, assuming that the clustering strength of the JWST AGN, like the type-1 AGN, is independent of luminosity \citep{Croom2005, Adelberger2006, Myers2006, Shen2009}, and we emphasise that this analysis does not need the intrinsic bolometric luminosity, which can be underestimated due to heavy obscuration (e.g. \citealp{Kocevski2024}) of the JWST AGNs.
Although no limits are set for luminosity, in the end, the AGNs selected by JWST are limited to those with lower luminosity ($-17<M_{UV}<-20$) than quasars.
Note that although we have selected JWST AGNs from literature with the same selection criteria described above, there may be a pre-selection of candidate sources in each spectroscopic survey.
Figure \ref{fig:sample_histogram} displays the redshift distribution of the JWST AGNs in the literature, showing that the number is the highest at $5<z<6$.
Hence, we select the JWST AGNs at $5<z<6$ (hatched region in Figure \ref{fig:sample_histogram}) for the clustering analysis.
We try detecting the clustering signal of the JWST AGNs at $z<5$ or $z>6$, but the signal is hardly detected due to their low surface number density at the redshift ranges.
The final sample contains the 27 JWST AGNs at $5<z<6$.

\subsection{Galaxies}
\label{subsec:galaxy}
We make use of the galaxy catalogue from DAWN JWST Archive (DJA)\footnote{\url{https://dawn-cph.github.io/dja/index.html}}.
DJA catalogues are created based on the public data of the JWST surveys, which are reduced with \texttt{grizli}\footnote{\url{https://github.com/gbrammer/grizli}} \citep{Brammer2023} and \texttt{msaexp}\footnote{\url{https://github.com/gbrammer/msaexp}} \citep{Brammer2022} by the Cosmic Dawn Center.
The catalogues contain photometric redshifts of the galaxies estimated by \texttt{EAZY}\footnote{\url{https://github.com/gbrammer/eazy-py}} \citep{Brammer2008} with JWST and Hubble Space Telescope (HST) photometry.
We use the v7 catalogues of three survey fields: Great Observatories Origins Deep Survey (GOODS: \citealp{Dickinson2003}) North and South; CEERS; PRIMER-UDS.
We note that the GOODS-North and GOODS-South catalogues contain the JADES and FRESCO data and that the RUBIES field is covered by the CEERS and PRIMER-UDS fields.

We select the bright galaxies from the catalogues by the following criteria:
\begin{align}
    & 5< z_\mathrm{phot} <6 \\
    &\&\,n_\mathrm{filter} \geq 12\\
    &\&\,23 < \mathrm{F444W} < 26, 
\end{align}
where $z_\mathrm{phot}$ is the photometric redshift by \texttt{EAZY} and $n_\mathrm{filter}$ represents the number of filters used to estimate the photometric redshift.
We use an aperture magnitude with a diameter of 0.\arcsec5.
We exclude the faint galaxies with $\mathrm{F444W}>26$ from the catalogue so that the depth of the limiting magnitude is uniform over the survey field.
As shown in Figure 2 of \citet{Merlin2024}, most of the fields show better sensitivity than 27 mag, which supports that the bright galaxies with $\mathrm{F444W}<26$ are homogeneously detectable.
We exclude extremely bright objects with $\mathrm{F444W}<23$ because some of them may be no extragalactic objects.
Homogeneous galaxy selection promises reliable cross-correlation analysis to derive the typical DMH mass of the JWST AGNs, although the JWST AGNs are not selected homogeneously.
Although we do not exclude the galaxies with poor EAZY template fitting, we confirm that the clustering strength does not change even when we limit the sample with the goodness of fit, $\chi_\nu^2<5$.
Finally, our sample contains the 679 galaxies that are distributed over 409.3 arcmin$^2$, and their breakdowns are summarized in Table \ref{tab:AGN_sample}.

\begin{table*}
	\centering
    \caption{The effective area and the number of the AGNs and the galaxies in each field. 
    }
	\begin{tabular}{lcccl} 
		\hline
		Field & Effective area & $N_\galaxy$ & $N_\AGN$ &  Reference\\
         & (arcmin$^2$) &  (\#) & (\#) &  \\
		\hline 
		GOODS North & 85.7 & 200 &  12  & \citet{Maiolino2023,Matthee2024}\\
		GOODS South & 62.1 & 69 &  2 & \citet{Maiolino2023,Matthee2024}\\
		CEERS & 95.3 & 207 & 9 & \citet{Harikane2023,Kocevski2024, Taylor2024}\\
        PRIMER-UDS & 166.2 & 203 & 4 & \citet{Kocevski2024, Taylor2024} \\
		\hline
            Total & 409.3 & 679 & 27 & \\
		\hline
	\end{tabular}\\
    \label{tab:AGN_sample}
\end{table*}

\section{Clustering Analysis} \label{sec:clustering_analysis}
We first evaluate the angular cross-correlation function $\omega(\theta)$ in Section \ref{subsec:angular} taking into account that the photometric redshift, whose uncertainty is larger than that of the spectroscopic redshift, is only available for the galaxy sample.
However, we note that the uncertainty of the photometric redshift is much smaller than the redshift range of the galaxy sample.
We also evaluate the projected correlation function $\omega_p(r_p)$ of each subsample in Section \ref{subsec:projected} to check the robustness of the result.
We note that these measurements of the typical DMH mass are independent.
We adopt almost the same way to evaluate the correlation functions in \citet{Arita2023}; therefore we briefly describe the method.

\subsection{Angular correlation function} \label{subsec:angular}
We evaluate the angular cross-correlation function between the JWST AGNs and the galaxies, $\omega_\mathrm{CCF}(\theta)$, and the angular auto-correlation function of the galaxies, $\omega_\mathrm{ACF}(\theta)$.
We use the following estimators to evaluate the correlation functions \citep{LS1993, Cooke2006}:
\begin{align}
    \omega_{\mathrm{CCF}}(\theta)&=\frac{D_\AGN D_\galaxy - D_\AGN R - D_\galaxy R + RR}{RR},\label{eq:ls_ccf_ang}\\
    \omega_{\mathrm{ACF}}(\theta)&=\frac{D_\galaxy D_\galaxy - 2D_\galaxy R + RR}{RR}, \label{eq:ls_acf_ang}
\end{align}
where $D_\AGN D_\galaxy, D_\AGN R, D_\galaxy R, RR, D_\galaxy D_\galaxy$ represent the normalized number of pairs between AGNs and galaxies, AGNs and random points, galaxies and random points, random points and random points, and galaxies and galaxies within the specified angular range, respectively.
The random points are scattered over the survey region at a surface number density of 100 arcmin$^{-2}$.
In order to trace the survey fields, we only use the random points located within $3\arcsec$ of the objects with $n_\mathrm{filter}\geq12$.
We evaluate both cross- and auto-correlation functions at $\theta>10\arcsec$ to avoid the one-halo term.
The uncertainties are estimated by the bootstrap resampling with $N=1000$ times iteration.
We randomly select the same number of the JWST AGNs and the galaxies from the sample allowing duplication and evaluate the cross- and auto-correlation functions for each subsample.
We calculate the covariance matrix below, and the diagonal element shows the uncertainty of each bin:
\begin{equation}
    C_{i,j} = \frac{1}{N-1}\sum_{k=1}^{N}(\omega_{i}^{k}-\overline{\omega}_i)(\omega_{j}^{k}-\overline{\omega}_j),
    \label{eq:covariance}
\end{equation}
where $\omega_{i}^{k}$ is the correlation function in $i$th bin of $k$th iteration and $\overline{\omega}_i$ shows the mean value of the correlation function in $i$th bin.

\begin{figure}
    \centering
    \includegraphics[width=\columnwidth]{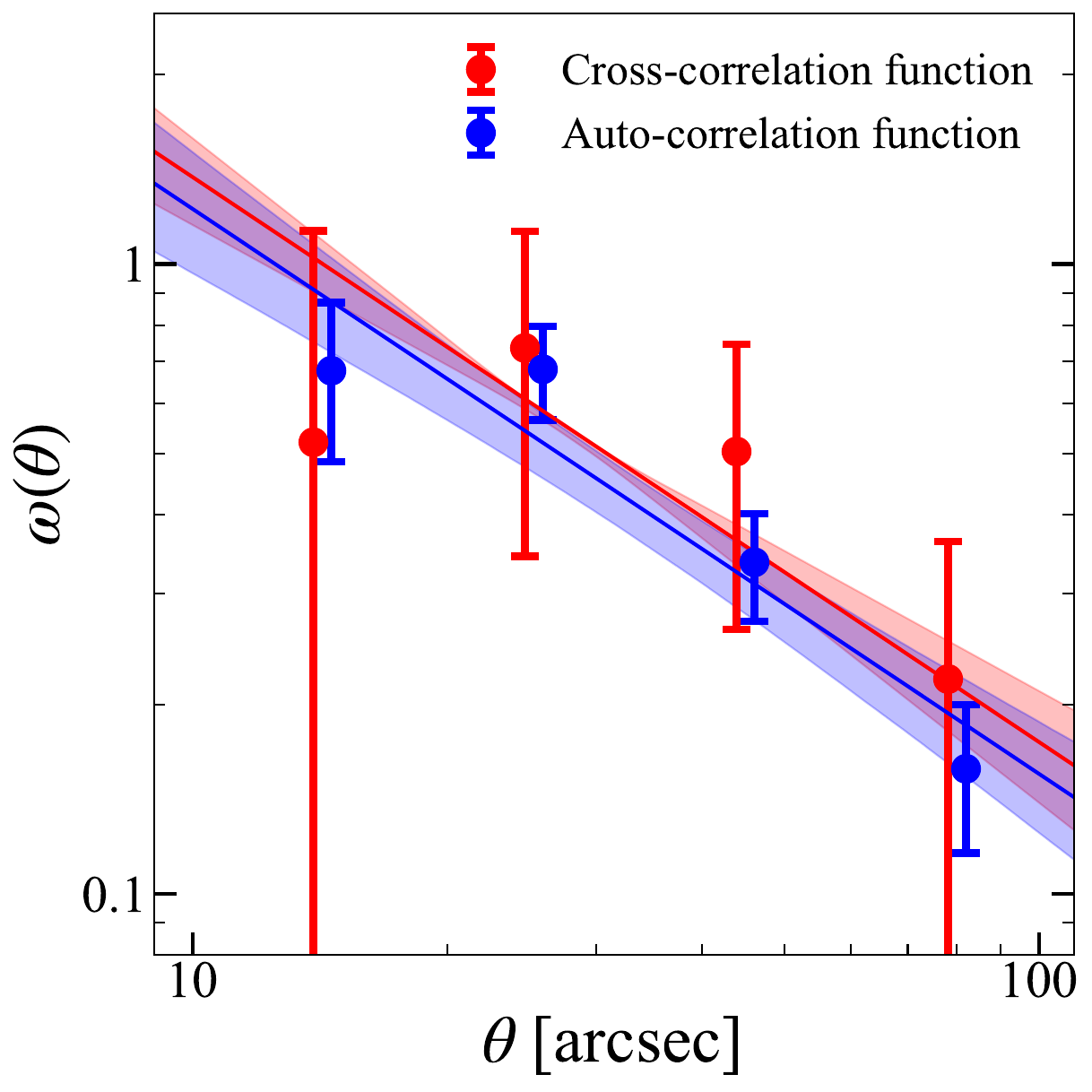}
    \caption{The angular correlation functions (blue: auto-correlation function; red: cross-correlation function).
    The data points show binned correlation functions, and the solid lines represent the best-fitting parametric correlation functions, in which we assume a power-law function $\omega(\theta)=(\theta/\theta_0)^{-\beta}$.
    The auto-correlation function signals are slightly offset in the $+x$-axis direction for visualization.
    The shaded regions denote the $1\sigma$ regions of the correlation functions.
    }
    \label{fig:angularCF}
\end{figure}

Figure \ref{fig:angularCF} shows $\omega_\mathrm{CCF}$ (red) and $\omega_\mathrm{ACF}$ (blue).
The red and blue solid lines show the best fit of a power-law function.
Regarding the integral constraint due to the limited survey area \citep{Groth1977}, we confirm that it can be negligible in a scale of $\theta\lesssim100\arcsec$.
Hence, we ignore the integral constraint in this analysis.

We use a Markov Chain Monte Carlo (MCMC) algorithm \citep{Foreman-Mackey2013} to fit the simple power-law function, $\omega(\theta)=(\theta/\theta_0)^{-\beta}$.
We assume a Gaussian likelihood function and uniform priors for $\theta_0\in[1\arcsec,100\arcsec]$ and the slope $\beta\in[0,2]$. 
We define the best estimate as the median and the 16th and 84th percentiles of the posterior distribution.
First, we perform the MCMC fit for the auto-correlation function because the signal-to-noise ratio is better than the cross-correlation function.
We obtain $\theta_{0,\ACF}=11.\arcsec93_{-2.81}^{+2.28}$ and $\beta=0.90_{-0.16}^{+0.16}$ as the best estimate.
The cross-correlation function uses the same $\beta$ obtained in the MCMC fit to the auto-correlation function, and $\theta_0$ for the fixed-$\beta$ is estimated in each MCMC step.
Finally, we obtain $\theta_0$ for the cross-correlation function as $\theta_{0,\CCF}=14.\arcsec26_{-2.12}^{+1.21}$.

The amplitude $A_\omega(=\theta_0^\beta)$ can be converted into the correlation length, $r_0$ in physical scale.
We calculate the correlation length of the cross-correlation function and the auto-correlation function by referring to \citet{Croom1999} and \citet{Limber1953}, respectively.
We use the redshift distribution by kernel density estimation with Gaussian kernel based on Figure \ref{fig:sample_histogram}.
Finally, we obtain $r_{0,\CCF}=6.33_{-0.71}^{+0.70}\,\hMpc$ and $r_{0,\ACF}=5.59_{-0.58}^{+0.59}\,\hMpc$.
In this analysis, we do not take the contamination fraction into account because it hardly affects the estimation for DMH mass measurement of the JWST AGNs.
The detail is described in Appendix \ref{appendix:contami}.

\subsection{Projected correlation function} \label{subsec:projected}
We also evaluate the projected cross-correlation function between the JWST AGNs and the galaxies, $\omega_{p,\mathrm{CCF}}(r_p)$ and the projected auto-correlation function of the galaxies, $\omega_{p,\mathrm{ACF}}(r_p)$.
The projected correlation functions are obtained by integrating the three-dimensional correlation functions, $\xi_\mathrm{CCF}(r_p,\pi)$ and $\xi_\mathrm{ACF}(r_p,\pi)$, where $r_p$ and $\pi$ represent the perpendicular and the parallel distances to the line-of-sight, respectively.
While the redshifts of the AGNs are determined spectroscopically, the galaxy sample only has the photometric redshifts.
It should be noted that the uncertainty of photometric redshift, typically $\Delta z/(1+z)=0.026$, is larger than that of spectroscopic redshift, and this may make the actual uncertainty of the correlation functions a little larger.
We adopt the following estimator to evaluate the three-dimensional correlation functions \citep{LS1993, Cooke2006}.
Namely, the projected correlation functions can be evaluated as
\begin{align}
    \omega_{p,\mathrm{CCF/ACF}}(r_p) = \int_0^{\pi_\mathrm{cut}} \xi_\mathrm{CCF/ACF}(r_p,\pi) d\pi
\end{align}
where $\pi_\mathrm{cut}$ represents the optimum limit above which the clustering signal is almost negligible, which is fixed to $100\,\hMpc$, and 
\begin{align}
    \xi_\mathrm{CCF}(r_p,\pi)=\nonumber\\
    \frac{D_\AGN D_\galaxy - D_\AGN R_\galaxy - D_\galaxy R_\AGN + R_\AGN R_\galaxy}{R_\AGN R_\galaxy}, \label{eq:ls_ccf_pro} \\ 
    \xi_\mathrm{ACF}(r_p,\pi) = \frac{D_\galaxy D_\galaxy - 2D_\galaxy R_\galaxy + R_\galaxy R_\galaxy}{R_\galaxy R_\galaxy}. \label{eq:ls_acf_pro}
\end{align}
In Equation (\ref{eq:ls_ccf_pro}) and (\ref{eq:ls_acf_pro}), $D$ and $R$ represent data and random points, respectively, and the suffixes denote the population.
The same random points in Section \ref{subsec:angular} are used, and their redshifts are assigned so as to reproduce the redshift distribution of the population shown in Figure \ref{fig:sample_histogram}.
Each term shows the normalized pair count within a specified projected length range.
The uncertainty of the projected correlation functions is estimated by the same method in Section \ref{subsec:angular}.
The covariance matrix is calculated by Equation (\ref{eq:covariance}) replacing $\omega$ for $\omega_p$.

\begin{figure}
    \centering
    \includegraphics[width=\columnwidth]{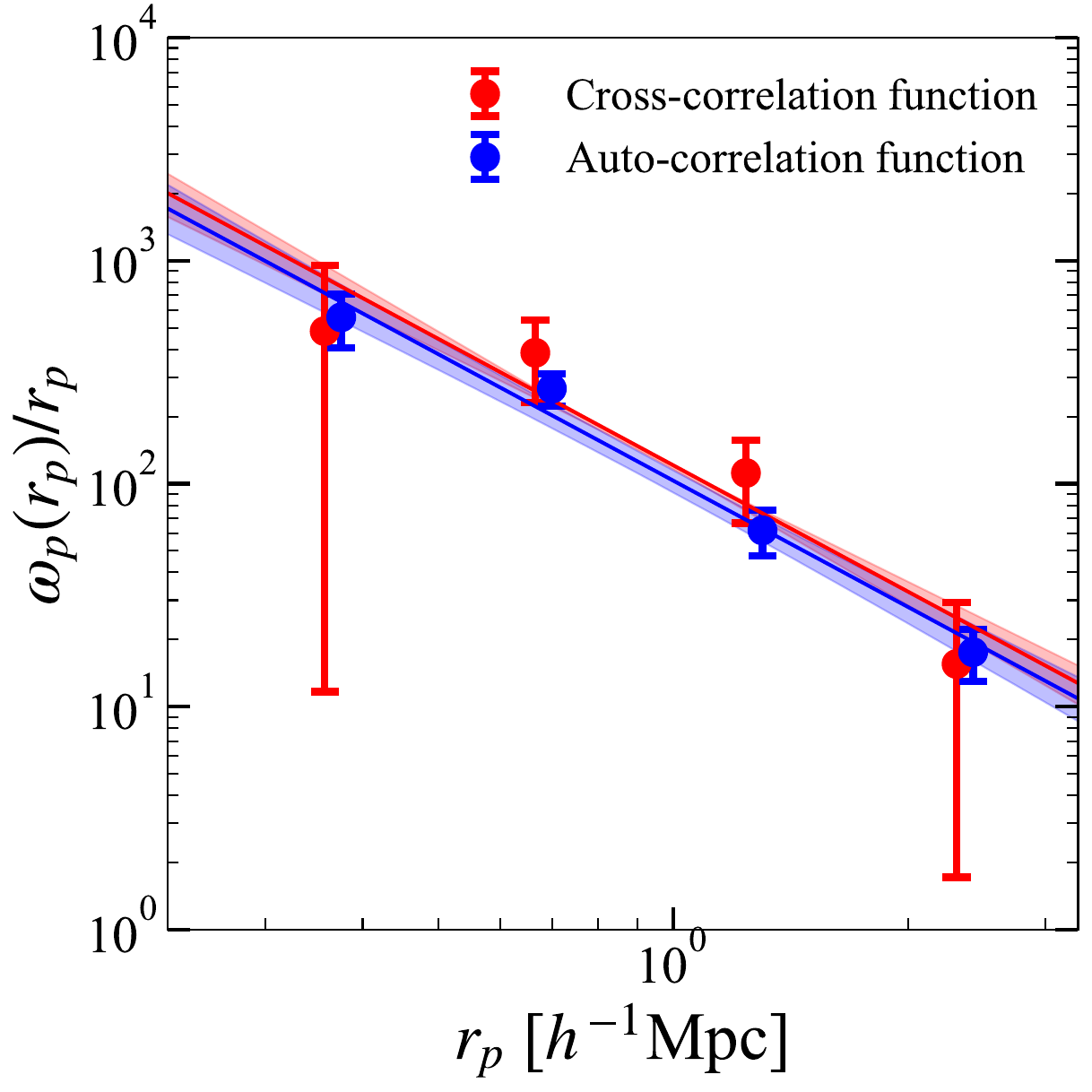}
    \caption{Same as Figure \ref{fig:angularCF}, but for the projected correlation functions.
    Here, we assume a power-law function of Equation (\ref{eq:omegap_th}) in the MCMC fit.
    }
    \label{fig:projectedCF}
\end{figure}

Figure \ref{fig:projectedCF} shows the result of the projected correlation functions.
The solid lines denote the power-law fit, which can be derived from the real-space correlation functions.
The relation between projected correlation functions and real-space correlation functions $\xi(r)$ is described in \citet{Davis1983} as
\begin{equation}
    \omega_p(r_p)=2\int_{r_p}^{\infty}\frac{r\xi(r)}{\sqrt{r^2-r_p^2}}dr.
\end{equation}
Assuming that the real-space correlation function is a power-law function expressed as $\xi(r)=(r/r_0)^{-\gamma}$, the projected correlation function is represented as
\begin{equation}
    \frac{\omega_p(r_p)}{r_p}=H_\gamma \left(\frac{r_p}{r_0}\right)^{-\gamma},
    \label{eq:omegap_th}
\end{equation}
where 
\begin{equation}
    H_\gamma = B\left(\frac{\gamma-1}{2}, \frac{1}{2}\right)
\end{equation}
in which $B$ is the beta function, and $r_0$ denotes the correlation length.
We fit Equation (\ref{eq:omegap_th}) to the projected correlation function with the MCMC algorithm.
Following \citet{Eilers2024}, we assume a Gaussian likelihood function and uniform priors for the correlation length of $r_0\in[1,30]\,\hMpc$ and the slope of $\gamma\in[1,3]$.
The best estimate is defined in the same manner in Section \ref{subsec:angular}.
Based on the MCMC fit for the auto-correlation function, we obtain $r_{0, \ACF}=5.80_{-0.60}^{+0.61}\,\hMpc$ and $\gamma=1.88_{-0.16}^{+0.16}$ as the best estimate.
The same $\gamma$ is used in the MCMC fit for the cross-correlation function to evaluate the correlation length.
Finally, we obtain the correlation length for the cross-correlation functions as $r_{0,\CCF}=6.63_{-0.67}^{+0.60}\,\hMpc$.

\subsection{DMH mass of the JWST AGNs} \label{subsec:DMHmass}
Based on the correlation lengths evaluated in Section \ref{subsec:angular} and \ref{subsec:projected}, we evaluate the typical DMH mass of the JWST AGNs.
We assume that the target objects are formed in the density peaks of the underlying dark matter and trace the peaks \citep{Sheth1999}.
The correlation length can be converted into a bias parameter, which is defined as the ratio of the 
clustering strength between the objects and the underlying dark matter at a scale of $8\,\hMpc$; therefore the bias parameter $b$ is obtained as
\begin{equation}
    b=\sqrt{\frac{\xi(8,z)}{\xi_\mathrm{DM}(8,z)}},
    \label{eq:def_bias}
\end{equation}
where $\xi_\mathrm{DM}$ represents the correlation function of the underlying dark matter.
We use \texttt{halomod}\footnote{\url{https://halomod.readthedocs.io/en/latest/index.html}}\citep{Murray2013,Murray2021} to calculate the denominator with the bias model of \citet{Tinker2010}, the transfer function model of \texttt{CAMB}\footnote{\url{https://camb.readthedocs.io/en/latest}} \citep{Lewis2011}, and the growth model of \citet{Carroll1992}.
With regard to the numerator, we assume that $\xi(r)=(r/r_0)^{-(1+\beta)}$ for the angular correlation function and $\xi(r)=(r/r_0)^{-\gamma}$ for the projected correlation function with the correlation lengths obtained in Section \ref{sec:clustering_analysis}.
We derive the bias parameter $b_{\mathrm{CCF}}$ and $b_\galaxy$ from the cross- and the auto-correlation functions with parameters in the MCMC steps, which is used to evaluate the uncertainty.
The bias parameters are summarized in Table \ref{tab:DMHmass}.
Finally, we estimate the bias parameter of the JWST AGNs $b_\AGN$ from \citet{Mountrichas2009}:
\begin{equation}
    b_{\mathrm{CCF}}^2 \sim b_\AGN b_\galaxy.
    \label{eq:bias_agn}
\end{equation}
This equation yields $b_\AGN=6.32_{-0.94}^{+0.89}$ and $6.61_{-0.82}^{+0.71}$ from the angular and the projected correlation functions, respectively.

We convert the bias parameters of the galaxies and the JWST AGNs into the typical DMH mass in the same method as \citet{Arita2023}.
Finally, through the results in Section \ref{subsec:angular}, we evaluate the typical DMH mass of the JWST AGNs and the galaxies as $\log (M_{\halo,\AGN}/\hMsun)=11.46_{-0.25}^{+0.19}$ and $\log(M_{\halo,\galaxy}/\hMsun)=11.12_{-0.22}^{+0.18}$, respectively.
Applying the results in Section \ref{subsec:projected} yields $\log(M_{\halo,\AGN}/\hMsun)=11.53_{-0.20}^{+0.15}$ and $\log(M_{\halo,\galaxy}/\hMsun)=11.18_{-0.22}^{+0.17}$, which are consistent with the results from the angular correlation functions.
In the evaluation of the typical DMH mass, we calculate the DMH masses for each bias from the parameters in the MCMC steps and regard the median and the 16th and 84th percentiles as the best estimate.
The results including bias parameters are summarized in Table \ref{tab:DMHmass}.

\begin{table*}
    \centering
    \caption{Summary of the clustering analysis results.
    All of the uncertainty in this table is defined as the median and the 16th and 84th percentiles based on the MCMC fit.}
    \begin{tabular}{lccccccc} \hline
         Correlation function & $r_{0,\ACF}$ & $r_{0,\CCF}$ & $b_\galaxy$ & $b_\CCF$ & $b_\AGN$ & $\log M_{\halo,\galaxy}$ & $\log M_{\halo,\AGN}$ \\ 
          & $(\hMpc)$ & $(\hMpc)$ & & & & $(\hMsun)$ & $(\hMsun)$ \\ \hline\hline
         Angular, $\omega(\theta)$ & $5.59_{-0.58}^{+0.59}$ & $6.33_{-0.71}^{+0.70}$ & $5.01_{-0.64}^{+0.59}$ & $5.64_{-0.74}^{+0.70}$ & $6.32_{-0.94}^{+0.89}$ & $11.12_{-0.22}^{+0.18}$ & $11.46_{-0.25}^{+0.19}$ \\
         Projected, $\omega_p(r_p)$ & $5.80_{-0.60}^{+0.61}$ & $6.63_{-0.67}^{+0.60}$ & $5.20_{-0.65}^{+0.60}$ & $5.89_{-0.68}^{+0.56}$ & $6.61_{-0.82}^{+0.71}$ & $11.18_{-0.22}^{+0.17}$ & $11.53_{-0.20}^{+0.15}$\\ \hline
    \end{tabular}
    \label{tab:DMHmass}
\end{table*}

\section{Discussion} \label{sec:discussion}
\subsection{Comparison of the typical DMH mass of the JWST AGNs and quasars} \label{subsec:redshift_DMHmass}
\begin{figure*}
    \centering
    \includegraphics[width=2\columnwidth]{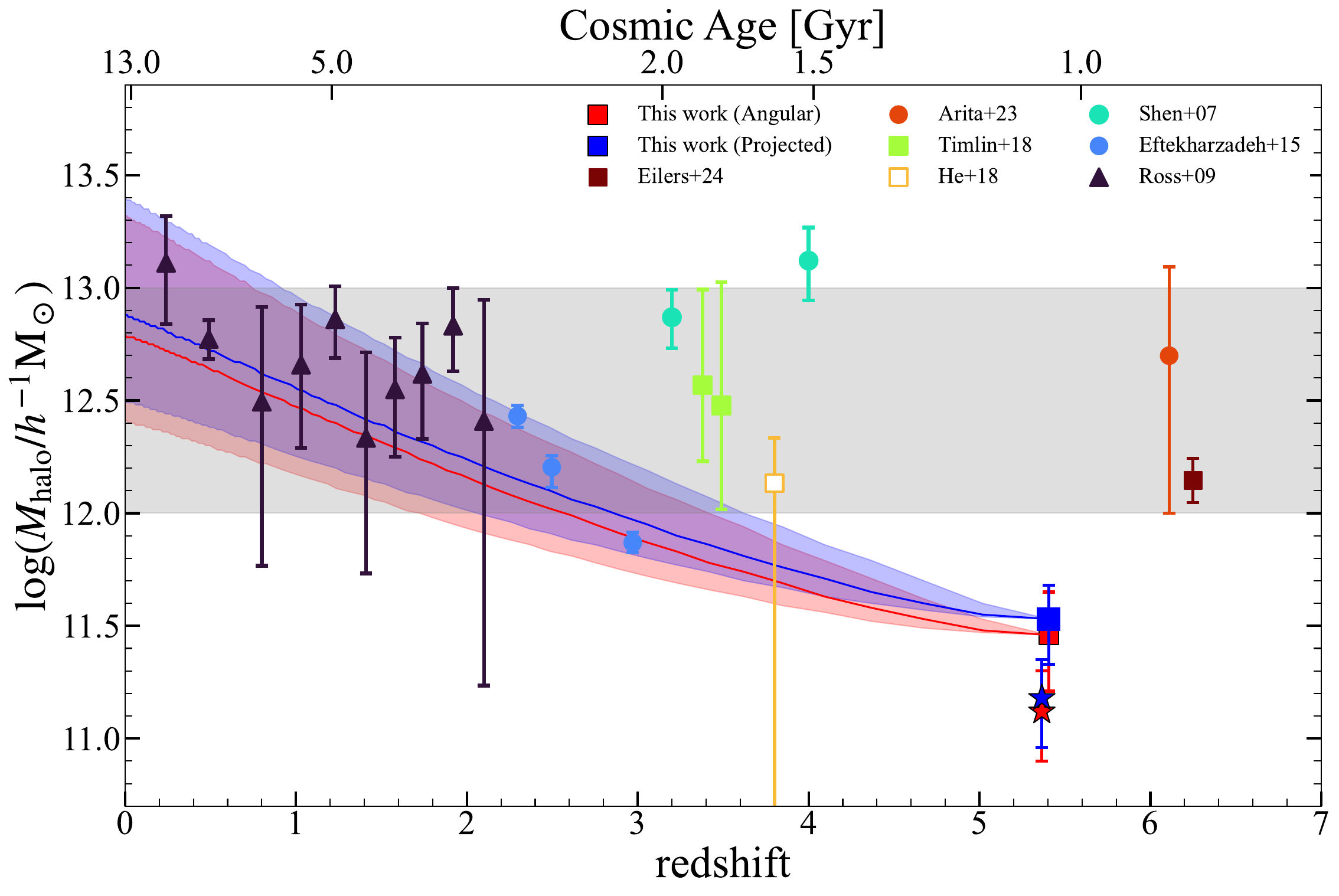}
    \caption{Comparison of the DMH mass of the JWST AGNs in this study (blue square: angular, red square: projected) with those in literature based on the clustering analysis.
    We also show the DMH mass of the galaxies as stars. 
    The symbols of the previous studies are classified by the type of the correlation function (circle: projected correlation function, square: angular correlation function, triangle: redshift-space correlation function).
    The filled and the open symbols show the auto-correlation function, and the cross-correlation function is used to estimate the typical DMH mass.
    We note that the DMH masses in the previous studies have been converted to those using the cosmological parameters in this study and their bias parameters because some cosmological parameters, particularly $\sigma_8$, have a large impact on the DMH mass estimate.
    The grey-shaded region shows the typical DMH mass range of quasars suggested by \citet{Trainor2012, Shen2013, Timlin2018, Arita2023}.
    The solid lines with the shaded regions denote the mass evolution of the DMH hosting the JWST AGNs at $5<z<6$ and its $1\sigma$ error calculated by the extended Press-Schechter theory. 
    }
    \label{fig:halo_mass_evolution}
\end{figure*}

We compare the typical DMH mass of the JWST AGNs with that of quasars derived by the clustering analysis.
Figure \ref{fig:halo_mass_evolution} summarizes the DMH mass measurement of quasars at $0\lesssim z\lesssim 6.5$ \citep{Shen2007, Ross2009, Eftekharzadeh2015, He2018, Timlin2018, Arita2023, Eilers2024}.
The previous clustering analysis indicates that type-1 quasars have a nearly constant halo mass of $\log(M_\halo/\hMsun) \sim12.5$ through the cosmic time \citep{Trainor2012, Shen2013, Timlin2018, Arita2023}.
\citet{Arita2023} discuss the possibility that there is a ubiquitous mechanism that activates quasars only in the DMHs with $12\lesssim\log(M_\halo/\hMsun)\lesssim 13$ (grey region in Figure \ref{fig:halo_mass_evolution}).
In contrast, the typical DMH mass of the JWST AGNs is lower than theirs by $\sim1$ dex, implying that the JWST AGNs are different populations from type-1 quasars.
\citet{Pizzati2024a} predicts that the DMH mass of LRDs should be smaller than that of unobscured quasars from their large abundance difference.
Although our sample is not necessarily identical to the LRDs, the DMH mass by the clustering analysis is consistent with the theoretical prediction.
No examples of DMH masses as less massive as the JWST AGNs in this study have been measured even in the faint type-1 quasars at low-$z$.
The DMH mass of the JWST AGN is rather consistent with that of the galaxy sample within 1$\sigma$ errors.
Given that they are different populations, there is no need for the abundance of the JWST AGNs on the LF to coincide with the type-1 quasar's extension to the faint-end, nor is there a need for the JWST AGN to follow the $M_\mathrm{BH}$-$M_*$ relation formed by the type-1 AGNs.
However, since the lower limit of the mass range of typical quasars has not been rigorously measured, faint quasars with $M_{1450}\gtrsim-20$ may reside in less massive DMHs with $\log(M_\mathrm{halo}/\hMsun)<12$.
Hence, the possibility that the JWST AGNs that are typically faint ($M_{1450}\gtrsim-20$) are new type-1 quasars hosted by less massive DMHs not previously found cannot be ruled out.

On the other hand, \citet{Allevato2014} reported that the DMH mass of X-ray-selected type-2 AGNs at $z\sim3$ is estimated as $\log (M_\halo/\hMsun)=11.73_{-0.45}^{+0.39}$, which is consistent with our measurements.
However, there are contradicting measurements of DMH mass for type-2 AGNs.
\citet{Allevato2011} showed that X-ray-selected narrow-line AGNs at $0.6\leq z\leq 1.5$ are hosted by massive DMHs with $\log (M_\halo/\hMsun)\sim 13.00\pm0.06$.
\citet{Viitanen2023} indicated that the DMH mass of X-ray-selected AGNs does not depend on their obsculation and that the typical DMH mass is $\log(M_\halo/\hMsun)=12.98_{-0.22}^{+0.17} (12.28_{-0.19}^{+0.13})$ at $z\sim0.7 (1.8)$.
The DMH mass of the JWST AGNs is less massive than that of X-ray selected AGNs \citep{Krishnan2020} at $0<z<2.5$, hosted on average in DMHs of $10^{12\mathchar`-13}\,\hMsun$.

We calculate the redshift evolution of the DMH mass of the JWST AGNs based on the extended Press-Schechter theory (e.g. \citealp{Bower1991}).
The red and blue solid lines in Figure \ref{fig:halo_mass_evolution} show the evolution of the DMH mass with $\log(M_\halo/\hMsun) = 11.46, 11.53$ at $z=5.4$, respectively.
We find that the DMHs hosting the JWST AGNs grow to as massive as $\sim 10^{13}\,\hMsun$ at $z=0$, which is comparable to the DMH mass of a galaxy cluster in the local Universe.
Furthermore, we find that the DMH mass of the JWST AGNs will reach $10^{12\mathchar`-13}\,\hMsun$, a typical type-1 quasar's DMH mass regime, at $z\lesssim 3$.
This recalls a scenario where the JWST AGNs at $5<z<6$ will grow into quasars at $z\lesssim 3$.
In other words, the JWST AGNs at $5<z<6$ are the progenitors of the quasars at $z\lesssim3$ and will start to shine as quasars in $\sim1$ Gyr later.
Here, from the perspective of the DMH mass evolution, it can be reasonably explained that the DMH hosting the JWST AGN will grow to be comparable to the DMH at $z<3$ quasar, but please note that this does not guarantee that the JWST AGN will necessarily grow to be a quasar. 
According to \citet{Hopkins2008}, an evolution model of quasars induced by a major merger, $\sim1$ Gyr before a quasar phase corresponds to a coalescence phase.
The model predicts that after the coalescence phase, a starburst occurs, significantly increasing the stellar mass of the host galaxy, and if this is correct, then at $z\sim3$, the overmassive situation in the JWST AGNs \citep{Maiolino2023, Harikane2023} is supposed to be mitigated.
The possibility of the episodic intense starburst for the JWST AGNs is remarked in \citet{kokorev2024b} based on their JWST observation for an LRD at $z=4.13$, which is consistent with the scenario.
The details on the relation between the host stellar mass and the SMBH mass are discussed in Section \ref{subsec:stellar_mass}.

Figure \ref{fig:halo_mass_evolution} also shows the typical DMH mass of our galaxy sample.
The DMH mass of the galaxies can also be inferred by combining their stellar mass, which is estimated by \texttt{EAZY} in the DJA catalogue, and the empirical stellar-to-halo mass ratio of \citet{Behroozi2019}.
Dividing the stellar mass of the individual galaxies by the stellar-to-halo mass ratio yields $\log(\overline{M}_{\halo,\galaxy}/\hMsun)=11.40_{-0.25}^{+0.26}$ as the median DMH mass of the galaxies, which is consistent with the DMH mass measured by the clustering analysis.
The consistency supports the robustness of the clustering analysis.

\subsection{Host stellar mass of the JWST AGNs} \label{subsec:stellar_mass}
The host stellar mass of the JWST AGNs can be inferred by the empirical relation between the stellar mass and the DMH mass of galaxies \citep{Behroozi2019}.
We multiply the typical DMH mass of the JWST AGNs by the stellar-to-halo mass ratio at $z\sim5.4$ to obtain the host stellar mass of the JWST AGNs as $\log (M_*/\mathrm{M_\odot}) = 9.48_{-0.41}^{+0.31}$ and $9.60_{-0.32}^{+0.24}$ from the angular and the projected correlation functions, respectively.
Our results are consistent with the stellar mass estimates for the LRD candidates at $5<z<6$ in the COSMOS-Web regions by SED fitting with the NIRCam and the MIRI photometry \citep{Akins2024}.
However, they are slightly higher than the estimate in \citet{Harikane2023} and $\gtrsim 1$ dex higher than the stellar mass estimated individually in \citet{Maiolino2023}.
\citet{Harikane2023} executed AGN decomposition based on the image before SED fitting, while \citet{Maiolino2023} performed SED fitting by \texttt{BEAGLE} \citep{Chevallard2016} with the AGN and the galaxy components to estimate the stellar mass.
\citet{Maiolino2023} remark that some of the estimated stellar masses are significantly smaller than the inferred dynamical mass, implying that their stellar mass might be underestimated due to the difficulty in assessing the AGN contribution in the spectra.
In addition, \citet{Casey2024} suggested that the stellar mass of LRDs is smaller than those obtained by the SED fitting when applying the maximum star-to-dust ratio \citep{Schneider2024}.
However, it remains to be elucidated whether the general star-to-dust ratio and the stellar-to-halo mass ratio can apply to the JWST AGNs.

\begin{figure}
    \centering
    \includegraphics[width=\columnwidth]{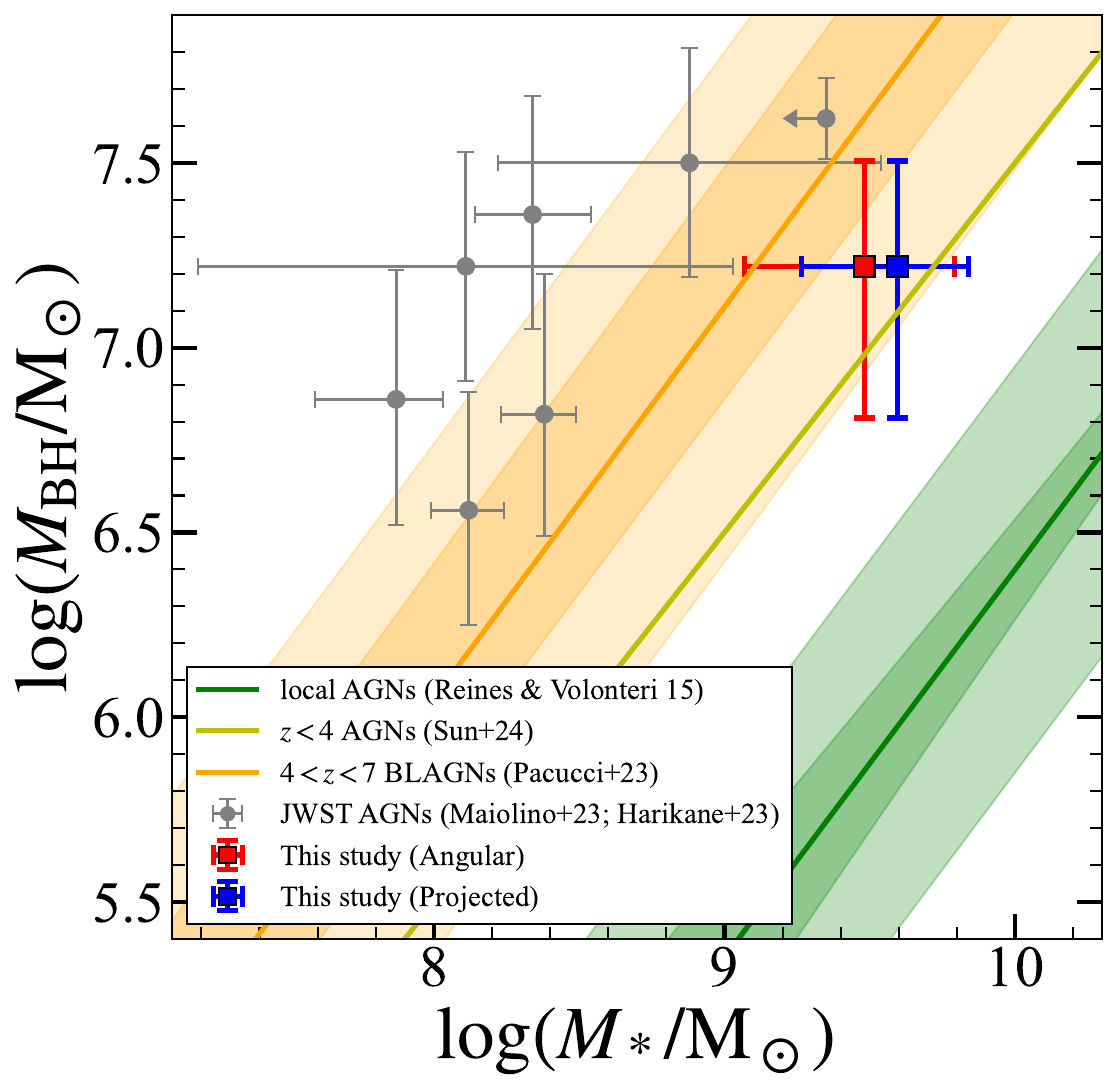}
    \caption{The relation between the stellar mass and the SMBH mass of the JWST AGNs. 
    The median stellar mass evaluated based on the DMH mass and the empirical stellar-to-halo mass ratio in this study are shown as red and blue squares.
    The grey points show the stellar mass and the SMBH mass of the individual JWST AGNs \citep{Maiolino2023, Harikane2023}.
    The green line denotes the relation of local AGNs \citep{Reines2015}.
    The orange line represents the relation based on the broad-line AGNs (BLAGNs) at $4<z<7$ identified by JWST \citep{Pacucci2023}.
    The deep and the light-shaded regions show the $1\sigma$ error and the intrinsic scatter, respectively. 
    The yellow line is the high-$z$ ($z<4$) relation suggested by \citet{Sun2024}.
    }
    \label{fig:coevolution_plot}
\end{figure}

Figure \ref{fig:coevolution_plot} compares the average $M_*$-$M_\mathrm{BH}$ relation for the JWST AGNs in this analysis and those in the literature.
The $M_\mathrm{BH}$ of the individual JWST AGNs has been measured based on the FWHM of the broad H$\alpha$ emission and the luminosity (e.g. \citealp{Greene2005, Reines2013, Reines2015}), and we use the median mass of the 27 JWST AGNs to evaluate their median $M_\mathrm{BH}/M_*$ in this study.
Figure \ref{fig:coevolution_plot} also displays the local relation \citep{Reines2015} and the high-$z$ relation \citep{Pacucci2023}, which is based on the JWST AGNs at $4<z<7$.
While \citet{Maiolino2023} and \citet{Harikane2023} reported that the JWST AGNs have highly overmassive SMBHs (grey circles in Figure \ref{fig:coevolution_plot}), our estimate shows the trend is much less pronounced.
Our results fall between the JWST AGNs' relation \citep{Pacucci2023} and the local AGN relation \citep{Reines2015}.
Our results are rather consistent with \citet{Sun2024}, who insist that $M_\mathrm{BH}/M_*$ shows no redshift evolution up to $z=4$.

\begin{figure}
    \centering
    \includegraphics[width=\columnwidth]{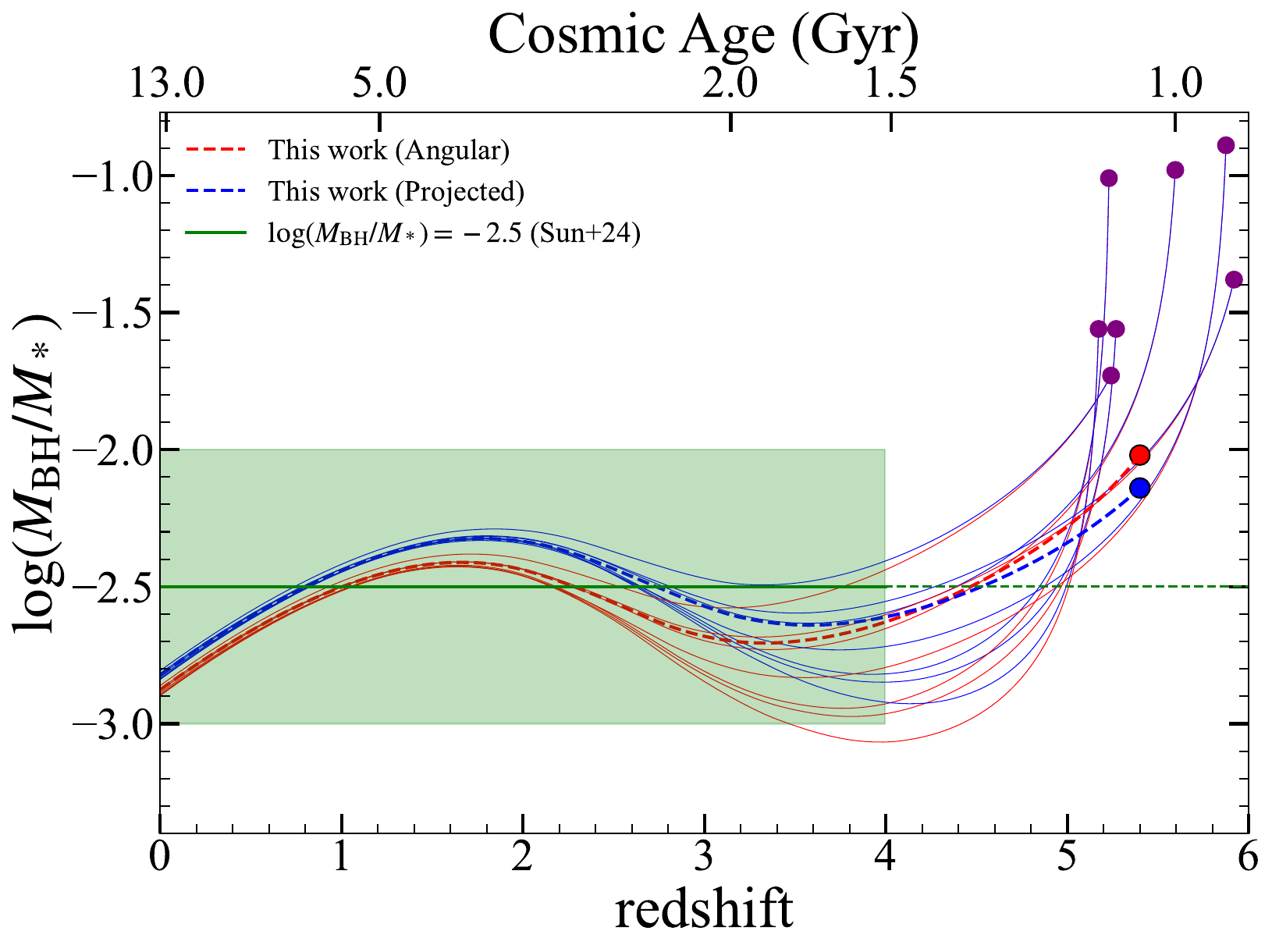}
    \caption{The redshift evolution of $M_\mathrm{BH}/M_*$.
    The red and blue dashed lines represent the evolution based on the angular and the projected correlation function.
    The solid lines show the evolution of the individual JWST AGNs whose stellar mass is estimated in the literature \citep{Maiolino2023, Harikane2023}.
    Their initial $M_\mathrm{BH}/M_*$ is represented as the purple circles.
    The red and blue ones calculate the BHAR based on the DMH mass evolution of the angular and the projected correlation functions.
    The green line denotes $M_\mathrm{BH}/M_*=-2.5$ at $z<4$ \citep{Sun2024}, and the green-shaded region displays a scatter of 0.5 dex.
    }
    \label{fig:MBHMstar}
\end{figure}

We also estimate a possible evolution of the $M_*$-$M_\mathrm{BH}$ relation for the JWST AGNs.
We adopt the black hole accretion rate (BHAR) of \texttt{TRINITY} \citep{TRINITY1}, which provides the average BHAR as a function of the redshift and the DMH mass.
\texttt{TRINITY} uses a halo merger tree and scaling relations among DMHs, galaxies, and SMBHs to predict the masses of DMHs, galaxies, and SMBHs at each redshift bin.
The BHAR can be obtained from the time derivative of the SMBH masses.
We simply assume that the host galaxies of the JWST AGNs have a constant star formation rate (SFR).
We estimate the total SFR by summing the SFRs derived by the H$\alpha$ luminosity in the narrow line component \citep{Matthee2024} and that from the IR luminosity \citep{Casey2024}.
\citet{Matthee2024} adopt \citet{Kennicutt2012} to estimate the SFR from the H$\alpha$ luminosity as $15\,\Msun\mathrm{yr^{-1}}$.
\citet{Casey2024} inferred the mean IR luminosity of the LRDs as $\langle L_\mathrm{IR}\rangle = (7.9_{-4.7}^{+2.9})\times 10^{10}\,\mathrm{L_\odot}$, which yields $\mathrm{SFR_{IR}}\sim 10\,\mathrm{\Msun\,yr^{-1}}$ by applying the empirical relation \citep{Kennicutt2012}.
Finally, we assume the SFR of the JWST AGNs as time-invariant with $25\,\mathrm{\Msun\,yr^{-1}}$.
Please note that there is a great amount of uncertainty involved in estimating the SFR.
If we stand on the model of \citet{Hopkins2008} mentioned in Section \ref{subsec:DMHmass}, the JWST AGN will undergo a starburst in the future, so this assumption may be underestimated.
Figure \ref{fig:MBHMstar} shows the evolution paths of $M_\mathrm{BH}/M_*$.
We confirm that most of the JWST AGNs will follow the low-$z$ relation \citep{Sun2024} at $z\lesssim3$, which supports the hypothesis that the JWST AGNs at $5<z<6$ are the progenitors of the quasars at $z\lesssim3$.
\texttt{TRINITY} predicts that AGNs residing in the DMHs with $M_{\halo,z=0}=10^{13}\,\hMsun$ will remain low in BHAR from $z\sim5$ to $z\sim3$, which helps mitigate the offset to the local relation at $z<3$.
In addition, \texttt{TRINITY} also predicts that the BHAR will start to increase at $z\sim3$, which implies that the AGNs become active at $z\sim3$. 
Furthermore, the recent JWST observation \citep{kokorev2024} for an LRD at $z=4.13$ shows a consistent $M_\mathrm{BH}$-$M_*$ relation with the local one \citep{Kormendy2013, Greene2016, Greene2020}.
They suggest that a starburst will occur after forming an overmassive SMBH at high-$z$, and the $M_\mathrm{BH}/M_*$ gets closer to the local relation.
Although this differs from the assumed star formation history, the final fate of the JWST AGNs is the same as the conclusion of this study.
The above assessment assumes that all of the JWST AGNs have an average DMH mass; however, the individual JWST AGNs should have different DMH masses and, therefore, different BHARs.
\citet{TRINITY6} also predicts the redshift evolution of $M_\mathrm{BH}/M_*$ of the JWST AGNs at $z\gtrsim 4$.
Their calculation shows that $M_\mathrm{BH}/M_*$ will keep almost constant or slightly increase toward $z=0$, which suggests that the JWST AGNs are still overmassive even at $z=0$, though no such population has been found.
The result may be because they assume mathematically that all SMBHs with the same host stellar mass share the same average Eddington ratio distribution. 
As such, the prediction of $M_\mathrm{BH}/M_*$ of the JWST AGNs has a large variation among the individual studies with different assumptions.

\subsection{Duty cycle} \label{subsec:dutycycle}
The duty cycle of AGNs, $f_\duty$, is defined as the time fraction of their active phase in the cosmic age.
Although numerous assumptions are needed to infer the duty cycle, we estimate it based on the DMHs of the JWST AGNs derived by the clustering analysis.
In order to estimate the duty cycle of the JWST AGNs, we assume that a DMH with $M_\mathrm{halo, min}\leq M_\halo\leq M_\mathrm{halo, max}$ hosts a JWST AGN and that JWST AGNs shines randomly in time.
Based on the assumption, the duty cycle can be derived as
\begin{equation}
    f_\duty = \frac{\int_{L_\mathrm{min}}^{L_\mathrm{max}} \Phi(L)\,dL}{\int_{M_\mathrm{halo, min}}^{M_\mathrm{halo, max}}n(M) dM},
\end{equation}
where $L_\mathrm{min}$ and $ L_\mathrm{max}$ are the minimum and maximum luminosity of the JWST AGNs used in the clustering analysis, respectively.
$\Phi(L)$ represents the LF of the JWST AGNs, and $n(M)$ denotes the DMH mass function.
Although it is under discussion which functions (e.g. double power-law, Schechter, or else) are appropriate to describe as the LF of the JWST AGNs, we refer to the LF derived in \citet{Matthee2024} and average the values at $M_\mathrm{UV}=-18.0, -19.0, -20.0$ to obtain $\log(\Phi(M_\mathrm{UV})/\mathrm{Mpc^{-3}\,mag^{-1}})=-4.94_{-0.22}^{+0.18}$.
We assume $M_\mathrm{halo,min}$ and $M_\mathrm{halo,max}$ as $10^{11}\,\hMsun$ and $10^{12}\,\hMsun$, respectively, which covers our DMH mass estimates.
The mass range is arbitrarily determined, and it is undeniable that the duty cycle significantly changes depending on how this range is taken.
It should be noted that even if we apply the usual definition of duty cycle (e.g. \citealp{Eftekharzadeh2015, He2018}) with $M_\mathrm{halo, max}$ replaced by $\infty$, the resulting value of $f_\duty$ does not change significantly.
We adopt the DMH mass function of \citet{Sheth1999}.
Finally, we obtain $f_\mathrm{duty}=0.37_{-0.15}^{+0.19}$ per cent, which implies the lifetime of $4\times10^6$ yr for the JWST AGNs at $5<z<6$, which is comparable to that of type-1 quasars at $z\lesssim3$ \citep{White2012, Eftekharzadeh2015, Laurent2017}.
This result also supports the scenario that the JWST AGNs are the progenitors of the quasars at $z\lesssim 3$.
If the DMH mass of the JWST AGN is as massive as that of quasars, namely $M_\mathrm{halo, min}=10^{12}\,\hMsun$, the duty cycle would be larger than unity.
This means that all of the massive DMHs host type-1 AGNs, and consequently this case does not allow inactive SMBHs and type-2 AGNs to reside in the massive DMHs at $5<z<6$.
Thus, it is qualitatively inferred that the JWST AGNs reside in less massive DMHs than quasars.
The duty cycle of $0.36_{-0.14}^{+0.18}$ per cent obtained here is self-consistent with the results that the typical DMH mass of the JWST AGNs is less massive than that of quasars, as long as the AGN activity is considered to be a transient phenomenon in the host galaxy.

\begin{figure*}
    \centering
    \includegraphics[width=2\columnwidth]{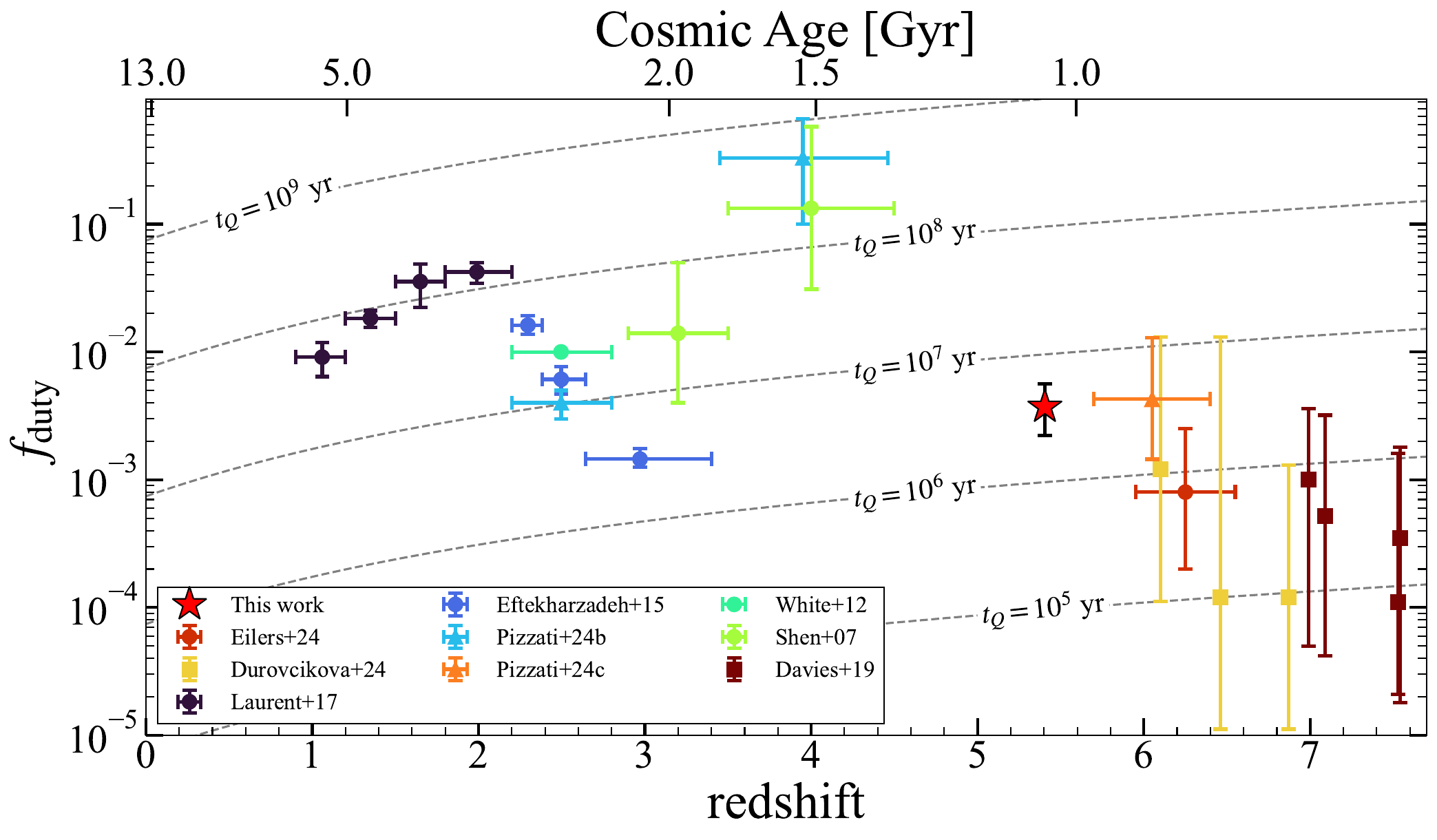}
    \caption{Duty cycle of quasars and the JWST AGNs.
    Our result is plotted as a red star.
    The markers of the previous studies show which method is adopted to estimate the duty cycle of quasars (circle: clustering analysis, square: Ly$\mathrm{\alpha}$ damping wing: triangle: cosmological simulation with clustering analysis).
    The dashed lines show $t_Q=10^5, 10^6, 10^7, 10^8, 10^9$ yr, where $t_Q$ represents the quasar lifetime.
    }
    \label{fig:duty_cycle}
\end{figure*}

Figure \ref{fig:duty_cycle} compares the duty cycle of the JWST AGNs $f_\mathrm{duty,\,JWST\,AGNs}$ with those of quasars $f_\mathrm{duty,\,quasars}$.
Some of the quasar duty cycles are inferred based on the clustering analysis by estimating the minimum DMH mass to host a quasar \citep{Shen2007, White2012, Eftekharzadeh2015, Laurent2017, Eilers2024}.
\citet{Pizzati2024b} and \citet{Pizzati2024c} also use the clustering analysis, and they evaluate the mass function of DMH hosting a quasar to measure the quasar duty cycle with a dark-matter-only simulation.
Other studies estimate the duty cycle based on the Ly$\alpha$ damping wings of the quasar spectra \citep{Davies2019, Durovcikova2024}.
We find that the duty cycle of the JWST AGNs is comparable to that of quasars at $z<5$ while it is slightly higher than that of quasars at $z>6$.

However, we caution that the estimated duty cycle of the JWST AGNs has a large uncertainty.
It is difficult to determine the mass range of the DMHs hosting the JWST AGNs, and their LF has a large variation among the literature.
These uncertainties are inevitable for the estimate of the duty cycle using the method adopted in this study.

\subsection{The Nature of JWST AGNs} \label{subsec:nature_agn}
We discuss possible interpretations of the JWST AGNs based on the DMH mass, the host stellar mass, and the duty cycle estimated in this study.
We suggest the four possibilities on the nature of the JWST AGNs: (a) progenitors of low-$z$ quasars; (b) a new AGN population; (c) low-DMH-mass type-1 quasars; (d) non-AGN objects.
\begin{enumerate}
    \item[(a)] \textbf{Progenitors of low-$z$ quasars}: As described in Section \ref{subsec:redshift_DMHmass}, DMHs that host the JWST AGNs will grow to $10^{12\mathchar`-13}\,\hMsun$, a typical DMH mass range of quasars \citep{Trainor2012, Shen2013, Timlin2018, Arita2023} at $z\lesssim 3$.
    This interpretation is compatible with the scenario suggested by \citet{Hopkins2008}.
    The JWST AGN sees a period of coalescence with a significant host stellar mass growth, after which the SMBH mass rapidly increases, and the AGN enters the quasar phase after $\sim1$ Gyr.
    The $M_\mathrm{BH}/M_*$, which is overmassive at $z\sim5.5$, is also found to become consistent with the local AGN value at $z<4$ \citep{Sun2024}.
    The JWST AGN does not need to continue to be active until it becomes a quasar, and this is consistent with the evaluation in this study that the duty cycle is less than unity.
    However, we caution that significant uncertainties about the evolution of the SFR and the BHAR are inevitable.
    The fact that most of the JWST AGNs are non-detectable in X-ray can be explained if we consider that their BHARs are already declining.
    In addition, if the JWST AGNs see a period of coalescence, their H$\alpha$ luminosity should be enhanced, which can easily explain the deviation from the relation between the X-ray luminosity and the H$\alpha$ luminosity \citep{Yue2024}.
    If the JWST AGNs are progenitors to quasars, then similar objects could be found at any epoch in the universe. 
    It would be interesting to find such objects in the low-$z$ universe (e.g. \citealp{Juodzbalis2024}).

    \item[(b)] \textbf{A new AGN population}: At $5<z<6$, the typical DMH mass of the JWST AGNs is $\sim1$ dex smaller than that of quasars although it has large uncertainty ($\sim0.5$ dex).
    This difference suggests that the JWST AGNs and quasars are distinct AGN populations.
    For instance, \citet{Maiolino2024} proposed that the X-ray emission of the JWST AGNs is intrinsically weak.
    They suggested a narrow-line Seyfert 1 with a high accretion rate and AGNs without hot coronas as possible scenarios.
    It is necessary to understand the detailed mechanism of the X-ray weakness and the lack of flux variability observed in the JWST AGNs.
    Since the JWST AGNs are considered to be a different population from what we call type-1 AGNs, their LFs do not need to be loosely connected to each other, and their contribution to the reionization should be considered independently based on the different escape fraction of ionizing photons, which means that the JWST AGNs' contribution is not necessarily the same as that of the type-1 AGNs' contribution. 

    \item[(c)] \textbf{Low-DMH-mass type-1 quasars}: This study shows that the DMH mass of the JWST AGN is smaller than that of typical type-1 quasars, but this does not completely rule out that the JWST AGN is a type-1 quasar.
    The typical DMH mass range of quasars is suggested to be constant $12\lesssim \log(M_\halo/\hMsun)\lesssim 13$ across most of the cosmic time, but this DMH mass range is not rigorously measured, and it is not surprising that lower-DMH-mass type-1 quasars exist.
    In fact, \citet{Pizzati2024c} predicts a broader distribution of host halo masses.
    However, it is curious that no such low-DMH-mass type-1 quasars have been found in the nearby universe.
    This scenario most easily explains the observed broad Balmer lines, but of course, it continues to suffer from the previously noted problems of X-ray weakness and the lack of flux variability in the JWST AGNs.

    \item[(d)] \textbf{Non-AGN objects}: Since the DMH mass of the JWST AGNs is found to deviate from those of typical type-1 AGNs and it is closer to that of bright galaxies, it is also viable that they are non-AGN objects.
    In this case, the calculation of the duty cycle in Section \ref{subsec:dutycycle} should be reconsidered.
    Some features remain to be elucidated to determine whether the JWST AGNs are classified as AGNs.
    One of the features is the lack of flux variability reported in \citet{Kokubo2024}.
    They suggested that the broad Balmer lines may not originate from broad line regions in the AGNs but from the ultrafast outflow or H\textsc{i} Raman scattering (but see \citealp{Juodzbalis2024}).
    Another feature is the weak X-ray emission \citep{Yue2024, Maiolino2024}.
    In addition, \citet{Baggen2024} suggested that some of the broad Balmer lines in LRDs do not need AGN contribution by reflecting the kinematics of the host galaxies.
    They find that stellar density in the centre of several LRDs is extremely high, and therefore the velocity dispersion is also large (see also \citealp{Guia2024}).
    Thus, we caution that a detailed observation is needed to conclude whether the broad line components originate from AGNs or not.
\end{enumerate}

\section{Summary}\label{sec:summary}
In this paper, we conduct the clustering analysis to evaluate the typical DMH mass of the low-luminosity AGNs newly identified by JWST. 
We compile the literature to select 27 AGNs at $5<z<6$ whose broad Balmer lines have been spectroscopically detected by JWST, and select the 679 galaxies in the same fields over $409.3\,\mathrm{arcmin^2}$ from a public galaxy catalogue in DJA.
The main results are summarized below.
\begin{enumerate}
    \item 
    The angular and the projected cross-correlation functions yield $\log (M_\halo/\hMsun)=11.46_{-0.25}^{+0.19}, 11.53_{-0.20}^{+0.15}$, respectively, which are $\sim 1$ dex smaller than the typical DMH mass of quasars at $0<z<6$ derived by the clustering analysis.
    \item The DMH mass evolution based on the extended Press-Schechter theory suggests that the DMHs of the JWST AGNs at $5<z<6$ will grow to  $10^{12\mathchar`-13}\,\hMsun$ at $z\lesssim3$, a typical DMH mass of quasars at that epoch.
    This result implies that the JWST AGNs are progenitors of the quasars at $z\lesssim3$.
    \item The host stellar mass of the JWST AGNs is evaluated as $\log(M_*/\Msun)=9.48_{-0.41}^{+0.31}$ and $9.60_{-0.33}^{+0.24}$ based on the measured DMH mass and the empirical stellar-to-halo mass ratio \citep{Behroozi2019}.
    The mass is consistent with the inferred stellar mass in \citet{Akins2024}, who performed SED fitting with the NIRCam and the  MIRI photometry, while it is $\sim 1$ dex higher than the estimate by SED fitting after decomposing the AGNs based on the image \citet{Harikane2023} and the spectra \citet{Maiolino2023}.
    \item Assuming that the JWST AGNs are the progenitors of $z<3$ quasars, it is deduced that the SMBH-overmassive JWST AGNs will experience a starburst later stage based on the model of \citet{Hopkins2008} and approach the local $M_\mathrm{BH}$-$M_*$ relation. 
    We calculate the possible evolution of $M_\mathrm{BH}/M_*$ assuming the BHAR of \texttt{TRINITY} and the constant SFR of $25\,\Msun\,\mathrm{yr}^{-1}$.
    We find that the JWST AGNs will become consistent with the local relation of \citet{Sun2024} at $z\lesssim3$ while they are overmassive at $5<z<6$.
    \item We evaluate the duty cycle assuming that a DMH with $11\leq \log(M_\halo/\hMsun)\leq 12$ can host a JWST AGN, and they shine in a certain period randomly.
    We obtain the duty cycle of the JWST AGNs as $f_\duty=0.36_{-0.14}^{+0.18}$ per cent, which corresponds to the lifetime of $\sim4\times10^{6}\,\mathrm{yr}$.
    The duty cycle is comparable to that of quasars at $z<4$, while it is $\sim1\mathchar`-2$ dex higher than that of quasars at $z\sim6$.
    \item Based on the DMH mass measured in this paper along with other observational properties, we argue the following four possibilities: (a) progenitors of quasars at $z\lesssim3$; (b) a different AGN population from type-1 AGNs.
    We cannot exclude the other possibilities that (c) JWST AGNs are merely low-mass type-1 quasars or (d) non-AGN objects.
\end{enumerate}

Future JWST observations with NIRSpec IFU will reveal the AGN-driven outflow and the chemical enrichment of the host galaxies, which will provide important hints for understanding the nature of the JWST AGNs.
However, JWST AGNs, especially LRDs, have weak emission at non-optical wavelengths, which implies that other approaches are important to understand their nature.
In this paper, we have shown that clustering analysis and the derived DMH mass provide an independent clue to the connection between the newly discovered JWST AGN and the previously known population.
\citet{Schindler2024} recently reported a cross-correlation analysis with six galaxies and an LRD at $z=7.3$ to estimate the minimum DMH mass of the LRD as $\log(M_\mathrm{halo,\,min}/\Msun)=12.3_{-0.8}^{+0.7}$.
A larger sample of the JWST AGNs with a uniform selection is required to extend the clustering analysis of this work for the JWST AGNs.
The observations will enable us to evaluate the three-dimensional correlation function and the auto-correlation function of the JWST AGNs, which will allow us to more precisely evaluate their DMH mass.
They will have an immense impact on our understanding of the JWST AGNs.

\section*{Acknowledgements}
We are thankful to the anonymous referee for comments and suggestions.
We appreciate useful suggestions by Joseph Hennawi, Kohei Inayoshi, Elia Pizzati, Jan-Torge Schindler, Daming Yang, and Takumi Tanaka.
We also thank Matthew Malkan, Tommaso Treu, Sofia Rojas-Ruiz, and Kazuhiro Shimasaku for a fruitful discussion.

JA is supported by the Japan Society for the Promotion of Science (JSPS) KAKENHI Grant Number JP24KJ0858 and International Graduate Program for Excellence in Earth-Space Science (IGPEES), a World-leading Innovative Graduate Study (WINGS) Program, the University of Tokyo.
NK was supported by the Japan Society for the Promotion of Science through Grant-in-Aid for Scientific Research 21H04490.
MO is supported by the Japan Society for the Promotion of Science (JSPS) KAKENHI grant No. JP24K22894.
YT was supported by Forefront Physics and Mathematics Program to Drive Transformation (FoPM), a World-leading Innovative Graduate Study (WINGS) Program, the University of Tokyo, and JSPS KAKENHI Grant Number JP23KJ0726.

The data products presented herein were retrieved from the Dawn JWST Archive (DJA). DJA is an initiative of the Cosmic Dawn Center (DAWN), which is funded by the Danish National Research Foundation under grant DNRF140.
\section*{Data Availability}

The JWST AGN data and galaxy data were obtained from the literature and DJA, both of which are open to the public.
The derived data generated in this research will be shared on reasonable requests to the corresponding author.



\bibliographystyle{mnras}
\bibliography{reference}




\appendix

\section{Correction of contamination} \label{appendix:contami}
While the JWST AGNs are detected spectroscopically, the galaxies are selected based on photometric redshift, which will cause contamination in the galaxy sample, and the rate should be taken into consideration.
We simply assume that the contaminating objects are randomly distributed over the survey area.
In this case, the amplitudes of the cross- and the auto-correlation functions can be corrected as
\begin{align}
    A'_{\omega,\mathrm{CCF}}&=\frac{A_{\omega,\mathrm{CCF}}}{1-f_c^{\galaxy}},
    \label{eq:cont_CCF}\\
    A'_{\omega,\mathrm{ACF}}&=\frac{A_{\omega,\mathrm{ACF}}}{(1-f_c^{\galaxy})^2},
    \label{eq:cont_ACF}
\end{align}
where $f_c^{\galaxy}$ is the fraction of the contaminating sources in the galaxy sample \citep{He2018}.
As shown in the Limber's equation \citep{Limber1953} and Equation (\ref{eq:def_bias}), we obtain the following relation between the bias parameter and the amplitude:
\begin{equation}
    b\propto r_0^{\gamma/2}\propto \sqrt{A}.
\end{equation}
Then, adopting the relation to Equation (\ref{eq:bias_agn}) and assuming that the contaminating sources have little effect on the photometric redshift distribution or the contamination fraction is small, the following relation is derived:
\begin{equation}
    b_\AGN\sim\frac{b_\mathrm{CCF}^2}{b_\galaxy}\propto \sqrt{\frac{A_{\omega,\mathrm{CCF}}^2}{A_{\omega,\mathrm{ACF}}}}.
\end{equation}
Combining the equation with Equation (\ref{eq:cont_CCF}) and (\ref{eq:cont_ACF}) yields
\begin{equation}
    \sqrt{\frac{A_{\omega,\mathrm{CCF}}^2}{A_{\omega,\mathrm{ACF}}}} = \sqrt{\frac{{A^{'2}_{\omega,\mathrm{CCF}}}}{{A}'_{\omega,\mathrm{ACF}}}},
\end{equation}
which demonstrates that the contamination fraction does not affect the bias parameter of the JWST AGNs.



\bsp	
\label{lastpage}
\end{document}